\documentclass[%
 reprint,
superscriptaddress,
 amsmath,amssymb,
 aps,
floatfix,
]{revtex4-1}

\usepackage{graphicx}% Include figure files
\usepackage{dcolumn}% Align table columns on decimal point
\usepackage{bm}% bold math
\usepackage{amsmath}
\usepackage{xcolor}
\usepackage{comment}
\begin{document}

\preprint{APS/123-QED}

\title{Active Q-switched X-Ray
Regenerative Amplifier Free-Electron Laser\\
}% 

\author{Jingyi Tang}
\author{Zhen Zhang}
\email{zzhang@slac.stanford.edu}
\author{Jenny Morgan}
\author{Erik Hemsing}

\author{Zhirong Huang}
\email{zrh@slac.stanford.edu}

\affiliation{SLAC National Accelerator Laboratory, Menlo Park, California 94025, USA}

\date{\today}% It is always \today, today,
             %  but any date may be explicitly specified

\begin{abstract}
Despite tremendous progress in x-ray free-electron laser (FEL) science over the last decade, future applications still demand fully coherent, stable x-rays that have not been demonstrated in existing X-ray FEL facilities. In this Letter, we describe an active \textit{Q}-switched x-ray regenerative amplifier FEL scheme to produce fully coherent, high-brightness, hard x rays at a high-repetition rate. By using simple electron-beam phase space manipulation, we show this scheme is flexible in controlling the x-ray cavity quality factor \textit{Q} and hence the output radiation. We report both theoretical and numerical studies on this scheme with a wide range of accelerator, x-ray cavity, and undulator parameters. 
%and demonstrate that it is much less demanding to the electron accelerator and X-ray cavity requirements than other cavity-based approaches.

\end{abstract}

%\keywords{Suggested keywords}%Use showkeys class option if keyword
                              %display desired
\maketitle

%\tableofcontents

%\section{Introduction}
%X-ray free electron lasers (FELs) have opened a new window for the exploration of atomic and molecular science at the angstrom-femtosecond length and time scales characteristic of these phenomena~\cite{Bostedt2016, XFELapp}. Self-amplified spontaneous emission (SASE) is the most widely used operation mode for these X-ray facilities. SASE starts from shot noise in the initial electron beam distribution and results in a chaotic spiky spectrum with rather limited temporal coherence~\cite{kondratenko1980generating,BONIFACIO1984, KJK_sase}. Generating fully coherent, stable and high-brightness X-rays will lead to significant advances in cutting-edge research including coherent imaging, nonlinear spectroscopy, coherent quantum control and X-ray quantum optics~\cite{min2011coherent,kowalewski2017simulating, bergmann2021using, aquila2015linac, szlachetko2016establishing, glover2012x, adams2013x}. Self-seeding techniques have been shown to increase the temporal coherence of X-ray FELs~\cite{geloni2011novel, amann2012demonstration, inoue2019generation, nam2021high}, although they have limited spectral purity and can suffer from large intensity fluctuations.

X-ray free electron lasers (FELs) have unprecedented peak brightness compared with storage ring light sources and have opened a new window for the exploration of atomic and molecular science at the angstrom-femtosecond length and timescales~\cite{Bostedt2016, XFELapp}. Nevertheless, the low-repetition rate of the early x-ray FELs has limited the scientific throughput. The current development of high-repetition rate x-ray FELs will produce 2 to 3 order of magnitude increase in average spectral brightness beyond diffraction-limited storage rings. They will provide qualitatively new capabilities needed for \textit{in situ} and operando studies of real-world materials, functioning assemblies, and biological systems\cite{schoenlein2016lcls}. In addition, a high-repetition-rate accelerator can distribute the high-power electron beams to an array of FEL beamlines, with each FEL specialized for a certain class of experiments but all operating at a relatively high repetition rate. This multiplexing capability will significantly broaden the scientific reach and increase the user access time.

For hard x-ray FELs, self-amplified spontaneous emission (SASE) is the most widely used operation mode. SASE starts from shot noise in the initial electron-beam distribution and results in a chaotic spiky spectrum with rather limited temporal coherence ~\cite{kondratenko1980generating,BONIFACIO1984, kim_huang_lindberg_2017}. Generating fully coherent, stable and high-brightness x rays will significantly increase the spectral photon density and hence advance science opportunities in high-resolution x-ray spectroscopy, single particle imaging, coherent quantum control and X-ray quantum optics {\cite{wollenweber2021high,kowalewski2017simulating, bergmann2021using, aquila2015linac, szlachetko2016establishing, glover2012x, adams2013x}. Hard x-ray Self-seeding (HXRSS) techniques have been shown to increase the temporal coherence of x-ray FELs ~\cite{geloni2011novel, amann2012demonstration, inoue2019generation, nam2021high}, although they have limited spectral purity and can suffer from large intensity fluctuations. Based on its superconducting accelerator, the European XFEL HXRSS \cite{liu2023cascaded} produced the highest average spectral flux of $\sim 3\times 10^{15}$ photons/eV/sec at 8-9 keV among all light sources. 

Cavity-based x-ray free electron lasers (CBXFELs) such as the x-ray regenerative amplifier FEL (XRAFEL) and the XFEL oscillator (XFELO) have been proposed to produce highly coherent and stable hard x-rays ~\cite{huang2006rafel,Kim2009rafel}, especially for high-repetition rate FELs. In these proposals, high-brightness electron beams generate x-ray pulses in an undulator embedded within an X-ray optical cavity. The cavity is composed of a few Bragg mirrors with high reflectivity and narrow bandwidth at hard X-ray wavelengths and recirculates an intense monochromatic radiation pulse to seed the next fresh electron beam for amplification. After repetitive interactions, the temporal coherence of the output radiation is increased drastically with highly reproducible X-ray pulses. This leads to another 2 to 3 orders of magnitude increase in average spectral brightness compared with SASE FELs ~\cite{ryan2011, weilunFEL2017, Marcus2020,huang2023ming} and the current HXRSS performance.

%Cavity-based X-ray Free electron lasers (CBXFELs) such as the X-ray regenerative amplifier FEL (XRAFEL) and the XFEL oscillator (XFELO) have been proposed to produce highly-coherent and stable hard X-rays~\cite{huang2006rafel,Kim2009rafel}. These radiation sources will be two to three orders of magnitude higher in both peak and average spectral brightness compared to SASE FELs~\cite{ryan2011, weilunFEL2017, Marcus2020}. In these proposals, high-brightness electron beams generate x-ray pulses in an undulator embedded  within an X-ray optical cavity. The cavity is composed of a few Bragg mirrors with high reflectivity and narrow bandwidth at hard X-ray wavelengths and recirculates an intense monochromatic radiation pulse to seed the next fresh electron beam. Relative to the initial SASE source, the temporal coherence of the output radiation is increased significantly with highly reproducible X-ray pulses from shot to shot. 

Like optical laser cavities, the outcoupling method is one of the most critical components for CBXFELs. Most outcoupling methods for CBXFELs require manipulation of the cavity optics, such as using a thin drumhead part of the crystal~\cite{weilunFEL2017, thindiamond2016}, intracavity gratings~\cite{grating1_Karvinen, grating2_Makita, grating3, grating4_Li} or splitters~\cite{Yuri2019output_splitter}, pin-hole diamond mirrors~\cite{Freund_2019_pinhole,Marcus2020} or diamond mirrors with doping~\cite{Krzywinski:2019abu_doping}. A recent large-scale x-ray cavity experiment~\cite{margraf2023low} showed efficient and stable storage of x rays over many round-trips without any FEL interaction and demonstrated key cavity components including the in-coupling-out-coupling grating. However, CBXFELs have stringent requirements on the quality, stability, and radiation resilience of the x-ray optics~\cite{thermal_PhysRevB.86.054301,reflectivity_Kolodziej:yi5052, reflectivity2_shvyd2010high,reflectivity3_shvyd2011near}. Outcoupling methods involving cavity optics manipulation, e.g. using pinholes or doping, may lead to degradation of the crystal quality, while thin crystals or gratings may be subject to various thermal distortions. 

Unlike optical lasers, the gain medium of CBXFELs is a relativistic electron beam. Thus, it is natural to consider outcoupling methods that rely on electron-beam manipulations. References.~\cite{huang2006rafel,Tang2022,desy} suggest a simple outcoupling mechanism that leaks the radiation produced by a short electron beam outside the narrow Bragg bandwidth even though the output spectrum will be somewhat compromised. References.~\cite{Mcarthur2018,Margraf:2019qpy} also explore the use of special electron optics to repoint the microbunched electron beam after the FEL interaction for the purpose of radiation outcoupling. In this Letter, we propose a new and simple scheme that uses a linearly chirped electron beam to achieve active \textit{Q} switching \cite{siegman1986lasers} and enables flexible outcoupling for an XRAFEL. During FEL amplification, an electron beam with an initial energy chirp will be slightly compressed or decompressed by the undulator dispersion, which leads to a blueshift or redshift of FEL microbunching wavelength. As a result, the spectrum of the intracavity radiation can be purposefully shifted out of the narrow reflection bandwidth of the Bragg mirror and transmitted. By manipulating the initial electron energy chirp produced by the linear accelerator, we can control the cavity quality factor \textit{Q} and hence the intra-cavity power buildup as well as the output radiation.

This electron-beam-based \textit{Q}-switching is not only important to keep the cavity optics simple and intact, but also essential for a practical operation of a steady-state XRAFEL. Continuous wave, superconducting linacs such as the LCLS-II High-Energy (LCLS-II HE) ~\cite{lcls2hecdr} and SHINE FEL~\cite{shine} will provide high-energy electron beams at about 1 MHz repetition rate. Previous proposals of CBXFELs including XFELO~\cite{ryan2011, weilunFEL2017} and XRAFEL~\cite{Marcus2020} that lack active \textit{Q}-switching require the full machine repetition rate to feed a cavity of 300 m in round-trip length. Because pulses are outcoupled on each round trip, such designs requires the full electron beam repetition rate and power. They are also not compatible with multiplexing a high-repetition rate FEL for multiple undulator beamlines. Reducing the CBXFEL repetition rate could relieve some of these issues, but at the expense of much larger cavity lengths or substantial increases in cavity loss due to multiple passes and turn-by-turn outcoupling. Our scheme solves these problems because it allows the radiation to circulate and/or be amplified in a low-loss crystal cavity (with high \textit{Q}) for multiple round-trips before the maximum cavity power is dumped by active \textit{Q} switching which is controlled upstream by the electron-beam chirp. We also demonstrate the flexibility of our scheme in supporting a more compact x-ray cavity and an XRAFEL with a much lower electron energy than what is typically required for hard x-ray FELs.

\begin{figure}[htb]
   \centering
   \includegraphics[width=\columnwidth]{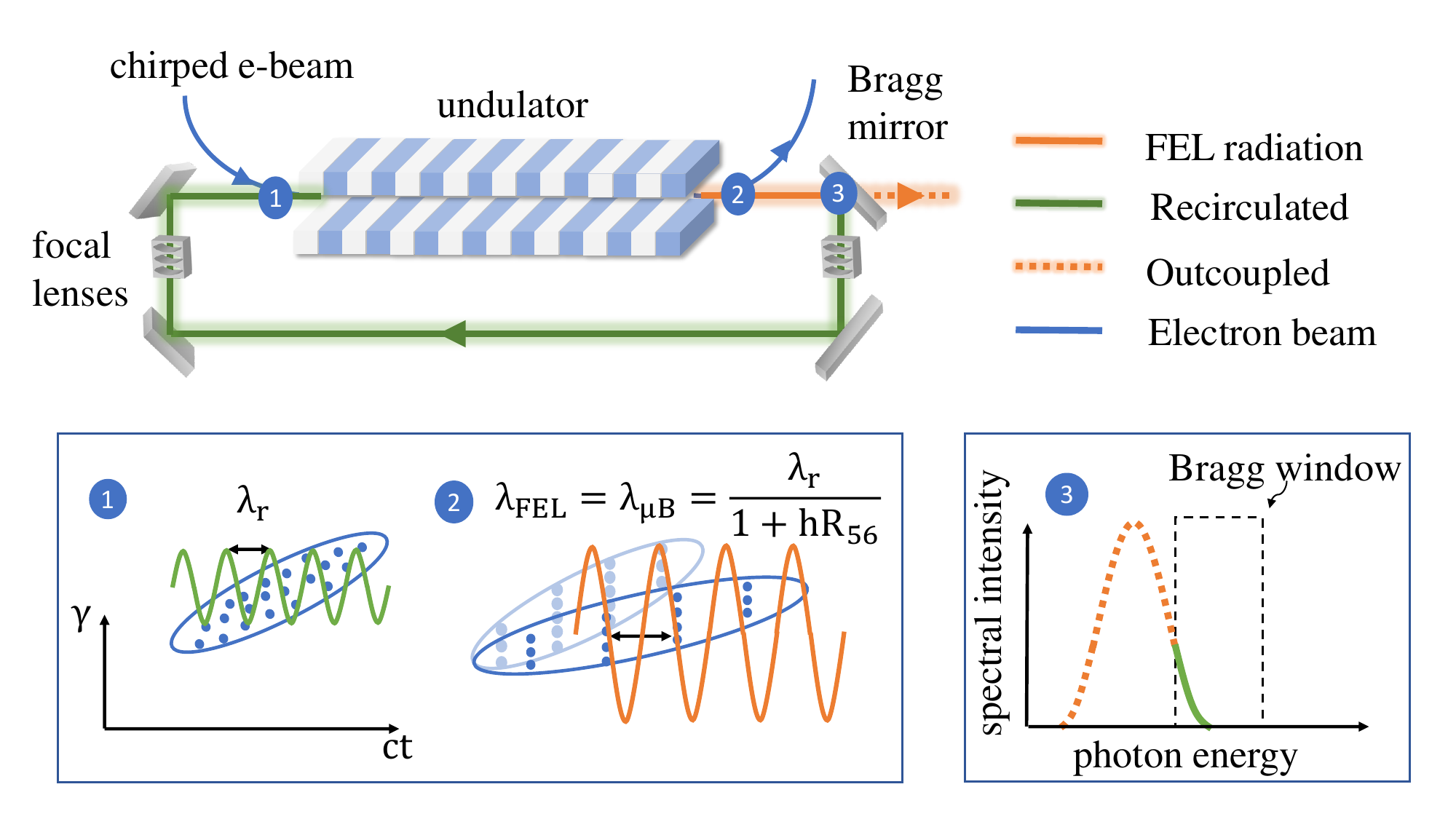}
   \caption{Illustration of an electron-beam-based \textit{Q}-switching scheme for an XRAFEL. Insets 1 and 2: electron longitudinal phase space evolution and FEL process. A chirped electron beam (blue line) will be slightly compressed or decompressed by the undulator (see text for details), resulting in a shift in the microbunching and the FEL wavelengths ($\lambda_{\mu \text{B}}$ and $\lambda_{\text{FEL}}$). Inset 3: the FEL radiation (orange line) spectrum is effectively switched out of the narrow reflection bandwidth of the Bragg mirror for output (orange dashed line). The remaining portion of the spectrum within the reflection window (green line) is recirculated to seed the next electron beam.}
   \label{fig:setup}
\end{figure}

% In this scheme, a moderate energy chirp is introduced to the electron beams to shift the spectrum of free electron laser (FEL) radiation outside the reflectivity bandwidth of the Bragg crystal. By actively controlling the chirp of the electron beam, the ratio of the out-coupled and recirculated pulse energy can be manipulated flexibly. This allows hard X-ray cavities driven by electron beams with reduced beam repetition rate, relatively low beam energy and short cavity length. In contrast to typical XRAFEL outcopling designs involving X-ray optics manipulation, this approach only requires the control of energy chirp of the electron beams, which can be simple and straightforward to implement.

% theory

We first illustrate the main physical mechanism here. The impact of an initial energy chirp on a seeded FEL has been analyzed in ~\cite{Lutman_2009,huang2010}. We consider an electron beam with a linear energy chirp given by
%For an electron beam with a central energy $\gamma_0 mc^2$ and a linear energy chirp 
\begin{equation}
    h=\frac{d\gamma/\gamma_0}{cdt}\,.
\end{equation}
The amplified seed signal in the high-gain presaturation regime and when the seed is much longer than  the FEL coherence length has a frequency shift given by
\begin{equation}\label{eq:freq_shift2}
    \frac{\Delta \omega}{\omega_r} = \frac{4}{3}  h\lambda_r N_u\,,
\end{equation}
where  $c$ is the speed of light, $m$ is the electron rest mass, $\gamma$ is the electron energy in units of $mc^2$, and $\gamma_0$ the average. $\lambda_r = \lambda_u(1+K^2/2)/(2\gamma_0^2) = 2\pi c/\omega_r$ is the FEL resonant wavelength, $\lambda_u$ is the undulator period, $K$ is the undulator strength parameter, and $N_u$ is the number of the undulator period. $\Delta \omega$ is the angular frequency shift relative to $\omega_r$. 

The result in Eq.~\eqref{eq:freq_shift2} can be understood from the view of bunch compression. As sketched in Fig.~\ref{fig:setup} with a chirped electron beam, both the bunch length and the microbunching wavelength will be compressed or decompressed by the momentum compaction of the undulator (denoted as $R_{56}$). The (de)compression factor of the electron beam is $C = 1/(1 + h R_{56}) \approx 1 - h R_{56}$, where $R_{56}=2 N_u\lambda_r$ after $N_u$ undulator periods. Thus, the relative change of microbunching wavelength due to the (de)compression is simply $hR_{56}$. Nevertheless, the amplified radiation temporal profile after $N_u$ undulator period will slip ahead of electrons by $N_u/3$ wavelengths determined by the radiation group velocity~\cite{kim_huang_lindberg_2017}, and its phase is mostly determined by the microbunched beam at the undulator distance $2N_u\lambda_u/3$. Thus $R_{56}=4N_u \lambda_r/3$ which yields Eq.~\eqref{eq:freq_shift2}. 3D FEL simulations also confirm this result as shown in Fig.~1 of the Supplemental Material~\cite{note1}.
%Besides, the energy chirp of the beam will cause a slight broadening of the FEL %bandwidth, while the time-bandwidth product of the pulse is still very close to %the Fourier-transform limit(details in Section I in the Supplementary materials).
%Fig.~\ref{fig:theory_vs_sim} (a) shows the spectral center shift as a function of initial energy chirp of the electron beam. The 3D FEL code GENESIS~\cite{REICHE1999243} is used to simulate seeded FEL process assuming an ideal 200\,MW initial seed and LCLS-II HE beam and undulator parameters as listed in Table~\ref{tab:table1} Case I. We linearly fit the spectrum center shift rate in the exponential growth regime and the results are consistent with theoretical predictions from Eq.~\ref{eq:freq_shift2}.  Fig.~\ref{fig:theory_vs_sim} (b) shows that the X-ray pulse is shorter with increasing energy chirp because the head and tail parts of the radiation are less amplified as their energies are further away from FEL resonant condition. The corresponding bandwidth is also increased but the time-bandwidth product is still very close to the Fourier-transform limit. 
Since the bandwidth of the Bragg mirrors for hard X-rays can be narrow (typically $< 100$ meV), only a moderate amount of energy chirp ($\sim$ 0.2\,MeV/fs) is required to shift the radiation spectrum outside the reflectivity window. Such a modest energy chirp can be easily generated by linear accelerators and does not degrade the FEL gain in the cavity. The output radiation bandwidth is slightly increased compared to the crystal bandwidth but the time-bandwidth product is close to Fourier transform limit.

\begin{comment}
\begin{figure}
   \centering
   \includegraphics*[width=1.0\columnwidth]{figure2.pdf}
   \caption{Spectral center shift and bandwidth broadening vs. linear e-beam energy chirp. (a) Spectral center shift rate as a function of initial e-beam chirp in the exponential gain regime. Theoretical prediction (blue line) is compared with 3D Genesis simulation (orange dots).  (b) Pulse length shortening (black) and bandwidth broadening (green) as a function of initial e-beam chirp at $z = 17m$ (near saturation). The fourier limit bandwidth of a Gaussian distribution is shown in the green dashed line. Simulations assume 200\, MW initial seed power. Beam and undulator parameters used are listed in Table~\ref{tab:table1} Case I. The vertical line indicates the e-beam chirp used in the Case I simulation.}
   \label{fig:theory_vs_sim}
\end{figure}
\end{comment}

%\section{Simulation}

We demonstrate the flexibility and robustness of this scheme in three different cases as listed in Table~\ref{tab:table1}. Case I and II are in the context of the LCLS-II HE,  with a beam  repetition rate at 100 kHz (one tenth the full accelerator repetition rate and beam power). The electron beam-energy is 8 GeV,  the undulator parameter $K = 1.657$, and the undulator period $\lambda_u = 2.6$\,cm. As a representative example, we consider a rectangular cavity composed of four diamond mirrors oriented at 45 degrees (see Fig.~\ref{fig:setup}), with Bragg resonance centered at 9.8 keV (Miller indices 400). Two compound refractive lenses are placed equidistant from each other  to establish a stable transverse mode inside the cavity. Such a cavity configuration is also chosen for the CBXFEL demonstration experiment at LCLS~\cite{marcus2019cavity} due to its simplicity and small transverse dimension. The field propagation in the drift spaces between the cavity optical components is modeled by the Fresnel equation. Each refractive lens is treated as a lossless parabolic phase mask in transverse space. The diamond mirrors are modeled using the dynamical theory of X-ray Bragg diffraction\cite{Yuri2012}. The FEL process is simulated by 3D FEL codes GENESIS~\cite{REICHE1999243} assuming ideal electron beams with Gaussian current profile. The simulations on the start-to-end electron beams from LCLS-II HE can be found in the Supplementary Materials~\cite{note1}.  

%TODO: add a table
\begin{table}[b]%The best place to locate the table environment is directly after its first reference in text
\caption{\label{tab:table1}%
Electron Beam and Cavity Parameters.
}
\begin{ruledtabular}
\begin{tabular}{lcccc}
\textrm{Parameter}&
\textrm{Case I}&
\textrm{II} &
\textrm{III} &
\textrm{Units}\\
\colrule
Beam energy, $E$ & 8.0 & 8.0 &  3.0& GeV \\
Beam repetition rate, & 0.1  & 0.1 & 1.0 & MHz \\

Cavity round-trip length, $L_c$ & 300 &100 &100 & m\\
Embedded undulator length, & 128 &46 &46 & m \\
Undulator strength \textit{K} & 1.657 &1.657 & 0.675 &  \\
Undulator period, $\lambda_u$ & 2.6 &2.6 & 1.0 & cm \\
Resonant photon energy, $\hbar \omega_r$ & 9.83 &9.83 &6.95 &keV\\
Beam peak current, $I$&  & 2& & kA \\
Beam rms duration & & 20&  &fs\\
Normalized emittance, $\gamma \epsilon_x$, $\gamma \epsilon_y$ &  & 0.3, 0.3& & $\mu$m \\
Uncorrelated energy spread &  &1.0 & & MeV\\
Beam energy chirp & 0.33 & 0.33 & 0.25 & MeV/fs \\

%Bragg mirror thickness &100& $\mu m$\\

\end{tabular}
\end{ruledtabular}
\end{table}

\begin{figure}[htb]
   \centering
   \includegraphics*[width=0.9\columnwidth]{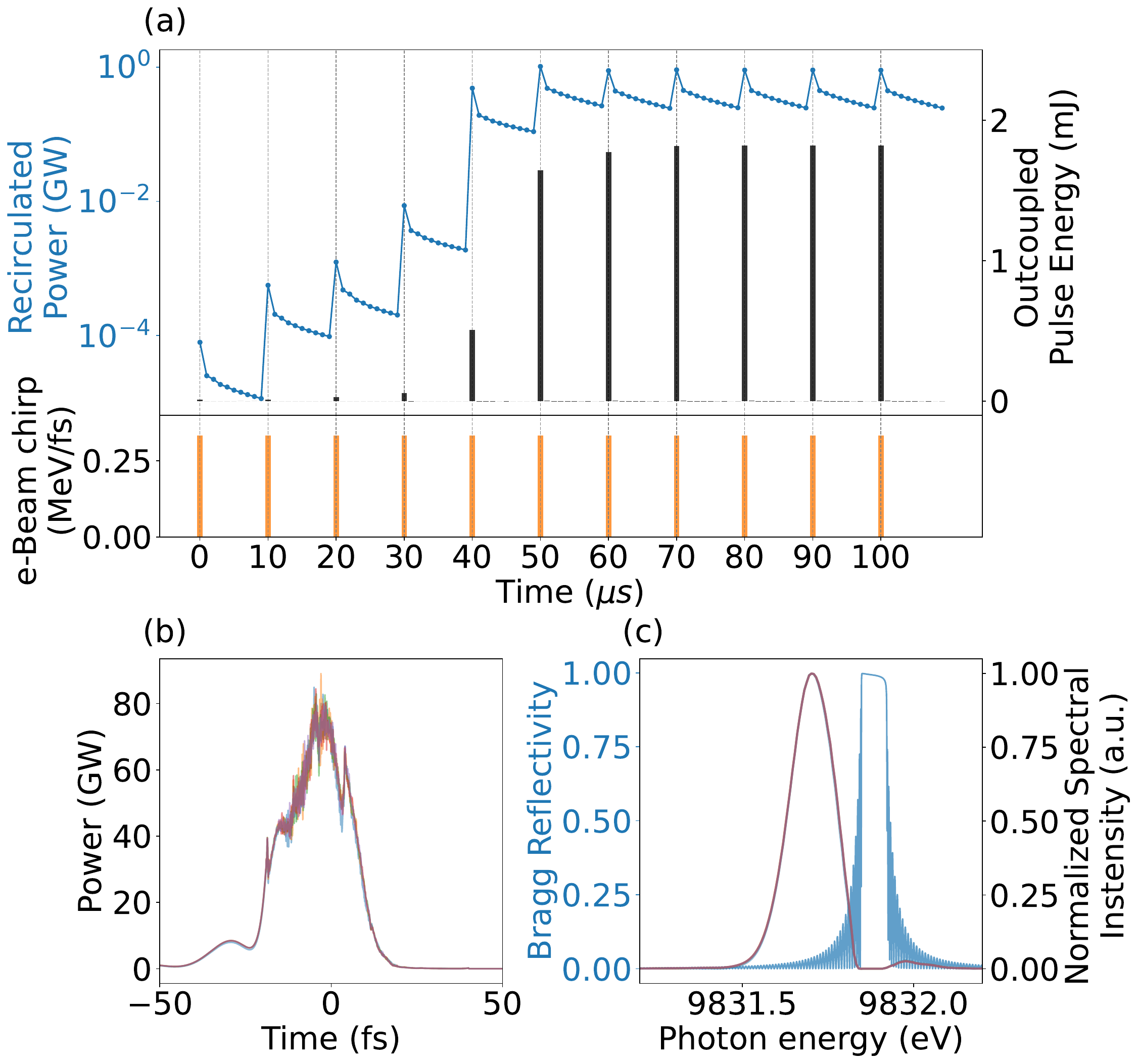}
   \caption{Case I: Electron-beam-based \textit{Q}-switching scheme with 100-kHz 8-GeV electron beam and 300-m cavity round-trip distance. (a) Recirculated and outcoupled power evolution as a function of time with FEL starting from shot noise. Each dot represents a recirculation pass in the cavity. Orange bars in the lower inset indicate the passes where a chirped electron is involved and amplifies the recirculated radiation.  (b) Five shots of outcoupled radiation power profile after reaching steady state. (c) Five shots of normalized on-axis spectra before (dashed line) and after (solid line) the first Bragg mirror. The blue curve represents the on-axis reflectivity of diamond (400). }
   \label{fig:long_cav}
\end{figure}

% TODO: more shots? longer beams? undulator wakefield? outcoupled power at passes without ebeam. 
% Q: better way to show brightness and compare with sase?

 In the configuration of LCLS-II HE, we first consider a cavity with round-trip distance $L_c = 300$\,m. 125-m undulators are embedded inside the cavity. The thickness of the diamond mirrors is $100~\mu m$ and the focal length of the lenses is $f = L_c/4 = 75 $\,m. With a 100-kHz electron beam rate, the radiation from the undulator recirculates in the cavity for ten round-trips before interacting with the next electron beam. Bragg mirrors act as a filter on the incident radiation in both frequency and transverse momentum space, resulting in 5-10\% of the pulse energy loss for each round trip in the cavity.  In this configuration, the long undulator line with optimized tapering can support high FEL gain, and only a small portion of the radiation energy is needed to build up a strong intracavity seed. As a result, we can use a chirped electron beam on every shot (i.e. at 100-kHz repetition rate) to \textit{Q} switch the cavity in order to dump out a significant fraction of radiation power. Figure~\ref{fig:long_cav}(a) shows the intracavity power buildup process starting from SASE with energy-chirped electron beams. The orange bars in the lower inset of Fig.~\ref{fig:long_cav}(a) represent the recirculation passes with a chirped electron beam, and the large FEL gain during these passes leads to a power spike in both the recirculated and outcoupled powers (blue and black curves). The intracavity power then decays slowly over the ten passes without an electron beam due to the high \textit{Q} of the cavity. By applying the proper amount of energy chirp on the beam, the center of the seeded FEL spectrum is shifted outside the Bragg reflection window and outcoupled from the cavity, while one of the spectrum tails remains inside the Bragg window to be recirculated and seed the next electron beam, as illustrated in Fig.~\ref{fig:long_cav} (c).  After the first five electron beams, the system reaches a steady-state condition with 1.75-mJ X-ray pulses outcoupled at the electron-beam rate of 100\,kHz. Figures ~\ref{fig:long_cav}(b) and (c) also show the five reproducible shots of outcoupled x-ray power profile and radiation spectrum. The FEL output power remains relatively stable to chirp jitter as the FEL is in the deep saturation regime. The x-ray stability due to various imperfections (such as energy chirp jitter and cavity misalignment) is not included in the simulation but is discussed in Secs. III and IV of the Supplemental Materials~\cite{note1, Qi2022}. The average spectral flux is on the order of $7.5\times 10^{17}$ photons/eV/sec, more than a factor of 100 higher than SASE at the same repetition rate or the demonstrated HXRSS at European XFEL\cite{liu2023cascaded}.

 %Both the peak and the average brightness of hard X-ray photons have been %improved by more than two orders of magnitude compared to SASE because of the %spectral narrowing. %The XRAFEL pulses are very reproducible. 

\begin{figure}[htb]
   \centering
   \includegraphics*[width=0.9\columnwidth]{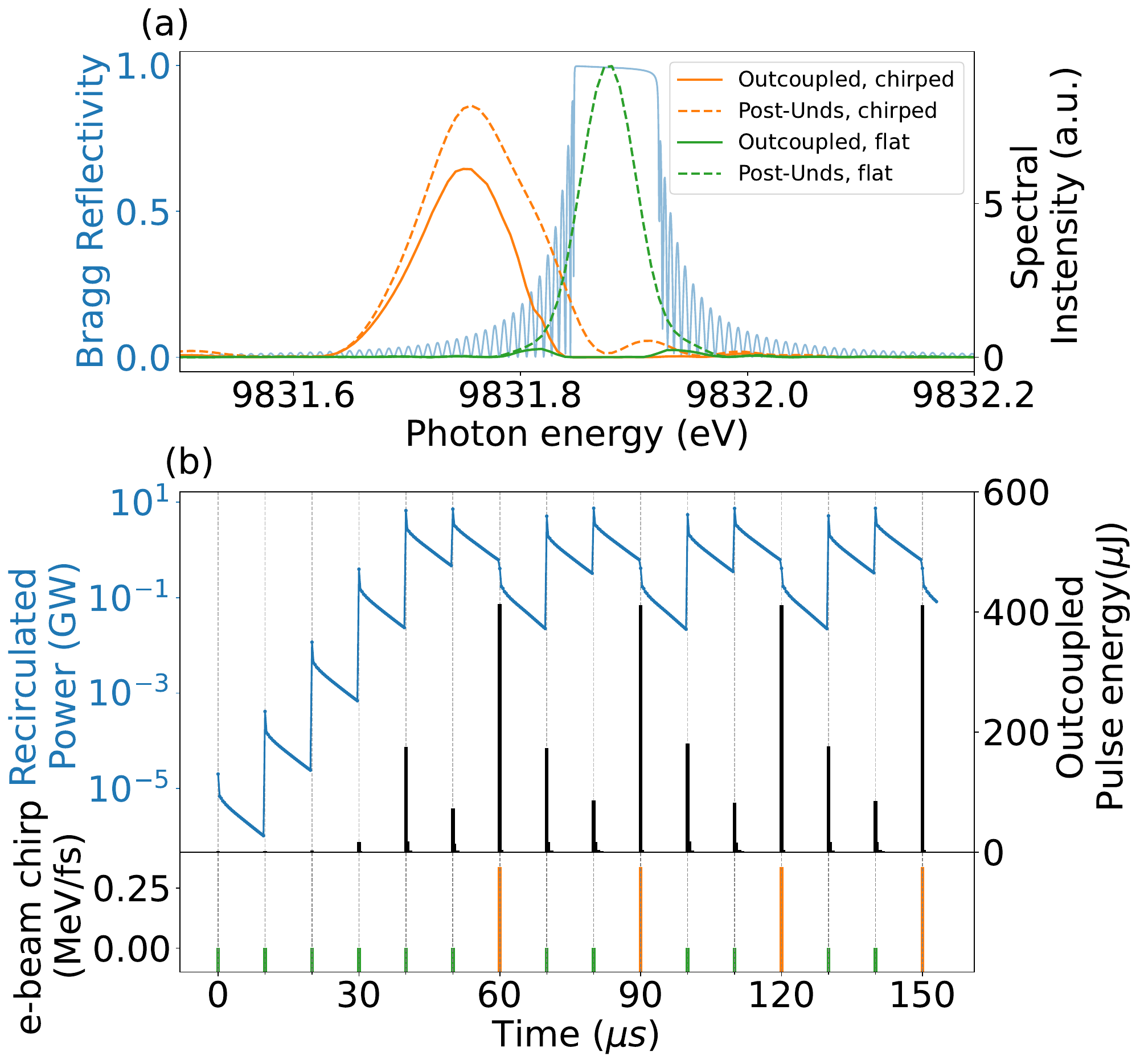}
   \caption{Case II: Electron-beam-based \textit{Q} switching scheme with 100-kHz 8-GeV electron beam and 100-m cavity round-trip distance. Flat and chirped electron beams are used alternatively to build the intracavity power and outcouple radiation. (a) On-axis spectrum of FEL radiation from flat electron beam (green) and chirped electron beam (orange). Solid curves are the spectra outcoupled from the first Bragg mirror while dashed curves are the spectra at the undulator exit. (b) Recirculated and outcoupled power evolution as a function of time with FEL starting from shot noise. Bars in the lower plot show the pass with electron beams, with green ones representing the flat electron beams and orange ones the chirped electron beams. }
   \label{fig:short_cav}
\end{figure}

Next we demonstrate that this \textit{Q}-switching scheme can be used to build a cavity with a more compact footprint, as listed in Table~\ref{tab:table1} Case II. This is achieved by alternating the electron energy chirp at a high repetition rate. Recently a short normal-conducting (NC) rf cavity with a fast filling time was proposed to control the electron compression and energy chirp on a shot-by-shot basis for LCLS-II~\cite{nasr2016cw,zhang2023fast}. Here we simulate a cavity with a 100-m round-trip distance and 46-m undulators embedded inside. In this case and with a 100-kHz electron-beam rate, the radiation is recirculated for 30 round trips before the next FEL amplification occurs. Thus, the radiation energy loss is substantial after 30 passes, and the FEL gain per pass is limited by the short undulator length. To maintain sufficient intracavity seed power, we can actively manipulate the electron energy chirp to control the cavity \textit{Q}.  As illustrated in Fig.~\ref{fig:short_cav}(a), flat electron beams (i.e., electrons without energy chirp) are used to build up intracavity power, and then chirped electron beams are used to outcouple the radiation. With flat electron beams, the center frequency of the FEL radiation remains in the Bragg reflection window and most of the radiation power will be recirculated. After the intracavity power reaches about 200\,MW with flat beams, a chirped electron beam is used to shift the spectrum outside the Bragg window and to dump the radiation power. The remaining intracavity power decreases significantly over the next 30 passes, only to be restored by a few more flat electron beams. Figure~\ref{fig:short_cav}(b) shows both the intracavity power evolution (blue) and the outcoupled radiation over time (black). Radiation from the flat beams can leak out of the Bragg bandwidth [see Fig.~\ref{fig:short_cav}(a)] and contribute to the output radiation. Nevertheless, its spectrum is spread out and can be filtered out by a postcavity monochromator centered at the shifted photon energy from the chirped beam.

\begin{figure}[htb]
   \centering
   \includegraphics*[width=0.9\columnwidth]{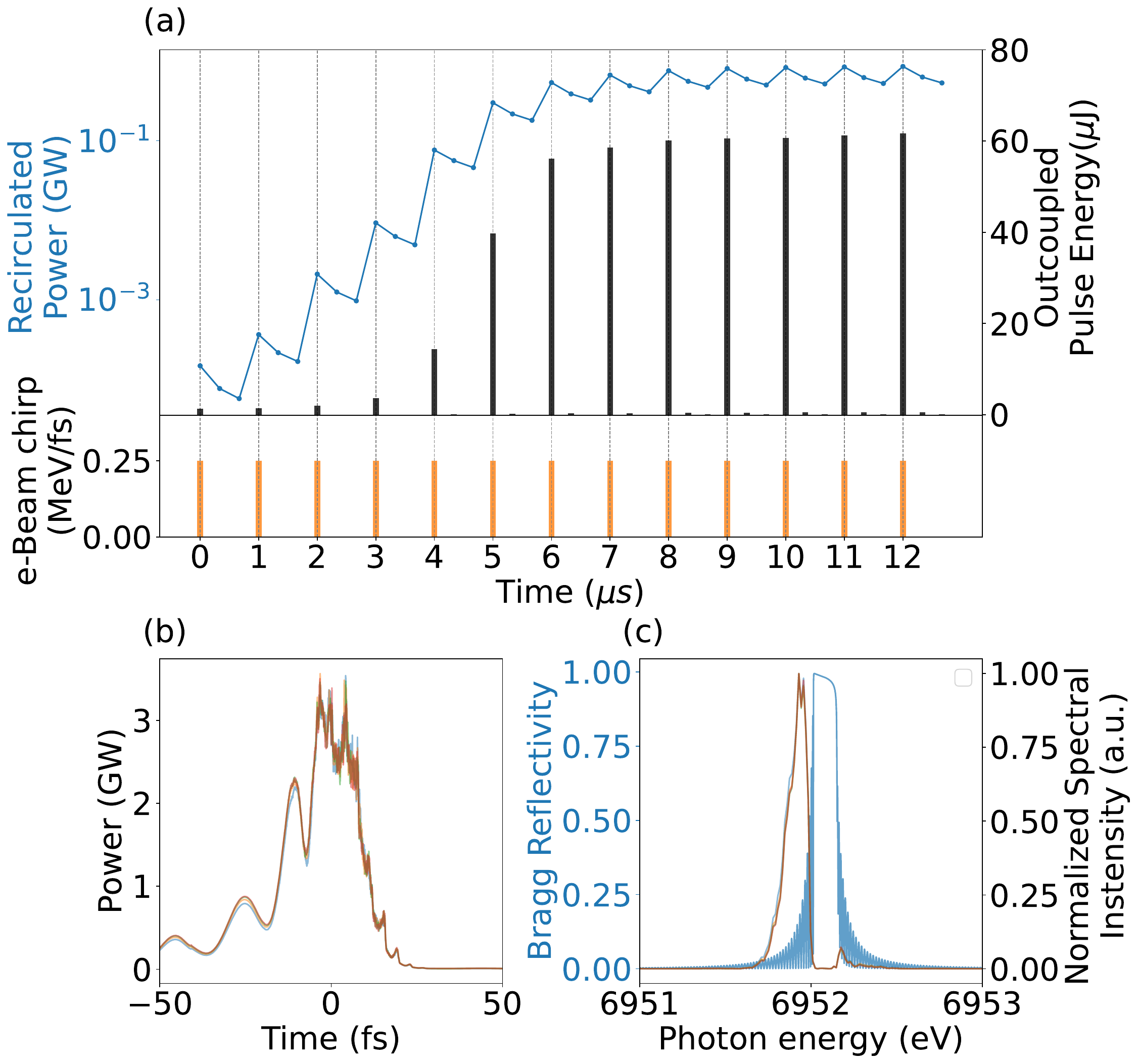}
   %\caption{Pass-to-pass FEL power evolution in the undulator starting from SASE  with 1MHz 3GeV chirped electron beam and 48 \,m undulators embedded in a cavity of 100m roundtrip distance. The resonant photon energy is 6.95 keV, which is the Bragg resonance of Diamond (220). }
   \caption{Case III: Electron-beam-based \textit{Q}-switching scheme with 1-MHz 3-GeV electron beam and 100-m cavity round-trip distance. (a) Recirculated and outcoupled power evolution as a function of time with FEL starting from shot noise. (b) Five shots of outcoupled radiation power profile after reaching steady state. (c) Five shots of normalized on-axis spectra of the outcoupled radiation. The blue curve represents the on-axis reflectivity of diamond (220).}
   \label{fig:3GeV}
\end{figure}

Finally in Case III of Table~\ref{tab:table1}, we explore a \textit{Q}-switched XRAFEL with a relatively low-energy superconducting linac designed for generating soft x-rays (e.g., UK XFEL~\cite{UKFEL} and Shenzhen XFEL~\cite{shenzhen-ipac23}). Together with the development of short-period, cryogenic permanent-magnet and superconducting undulators~\cite{BAHRDT2018149}, this can lead to significant cost savings for future x-ray FEL facilities. Figure~\ref{fig:3GeV} (a) shows the pass-to-pass power evolution of an XRAFEL driven by 3-GeV chirped electron beams at a 1-MHz repetition rate. 48-m undulators with $\lambda_u = 1$\,cm and $K = 0.675$ are embedded in the $L_c = 100$~m cavity. Miller indices (220) with resonant photon energy $\hbar \omega_r = 6.95$ keV are used as Bragg mirrors. A more conventional FEL undulator can also be used by having one of its higher harmonics tuned to the Bragg condition~\cite{harmonic}. %which is feasible given recent progress in superconducting undulators\cite{}. %Note that it is also possible to set the third harmonic  of FEL at the Bragg resonance center, which can allow the use of a normal undulator period. For example, in our case one can use $K = 0.915$ and $\lambda_u = 2.6$\,cm to make the cavity resonant at the third harmonics of FEL.  
If the thickness of the first Bragg mirror is $50~\mu m$ to reduce the absorption loss at this photon energy, 60-$\mu$J stable hard x-ray pulses with a narrow bandwidth can be outcoupled as indicated in Fig.~\ref{fig:3GeV}(b)-(c). Similar to Case II, we can reduce the repetition rate of the 3-GeV beam (down to 100 kHz) by alternatively feeding the cavity with the flat beams to build up sufficient intracavity power and with the chirped beams to dump out the power.

%The tolerance of cavity misalignment under this configuration may become a concern, as there are multiple roundtrips of X-rays in the cavity before seeding the electron beams.  Ref.\cite{Qi2022} shows that with the presence of the refractive lenses in the cavity, the misalignment-induced trajectory errors will oscillate rather than accumulate from pass to pass. Especially, in a confocal cavity as simulated in this paper, the spatial and angular deviation from the mirror misalignment will be canceled after an even number of recirculation passes. Besides, due to the high FEL gain, XRAFEL is much more resilient to such misalignment errors compared with XFELO. The numerical studies on the misalgnment tolerance shows robust power buildup with a few hundred nrad rms angular error on the mirrors, which is easy to obtain with conventional optomechanical hardware.

The authors would like to thank William M. Fawley, David Fritz, Alex Halavanau, Haoyuan Li, Gabriel Marcus, Robert Schoenlein and Diling Zhu for many useful discussions. This work was supported by the U.S. Department of Energy (Contract No. DE-AC02- 76SF00515) and Award No. 2021-SLAC-100732. 

% Acknowledge

%\begin{acknowledgments}

%\dots.
%\end{acknowledgments}

\appendix

%\bibliographystyle{IEEEtran}
%\bibliography{ref}% Produces the bibliography via BibTeX.

% Generated by IEEEtran.bst, version: 1.14 (2015/08/26)

\end{document}

\preprint{APS/123-QED}

\title{Supplimentary Materials: An Active Q-switched X-ray
Regenerative Amplifier Free-Electron Laser\\
}% 

\author{Jingyi Tang}
\author{Zhen Zhang}
\email{zzhang@slac.stanford.edu}
\author{Jenny Morgan}
\author{Erik Hemsing}

\author{Zhirong Huang}
\email{zrh@slac.stanford.edu}

\affiliation{SLAC National Accelerator Laboratory, Menlo Park, CA 94025, USA}

\date{\today}% It is always \today, today,
             %  but any date may be explicitly specified
   
\maketitle

\section{Theoretical Analysis}

\begin{figure}
   \centering
   \includegraphics*[width=1.0\columnwidth]{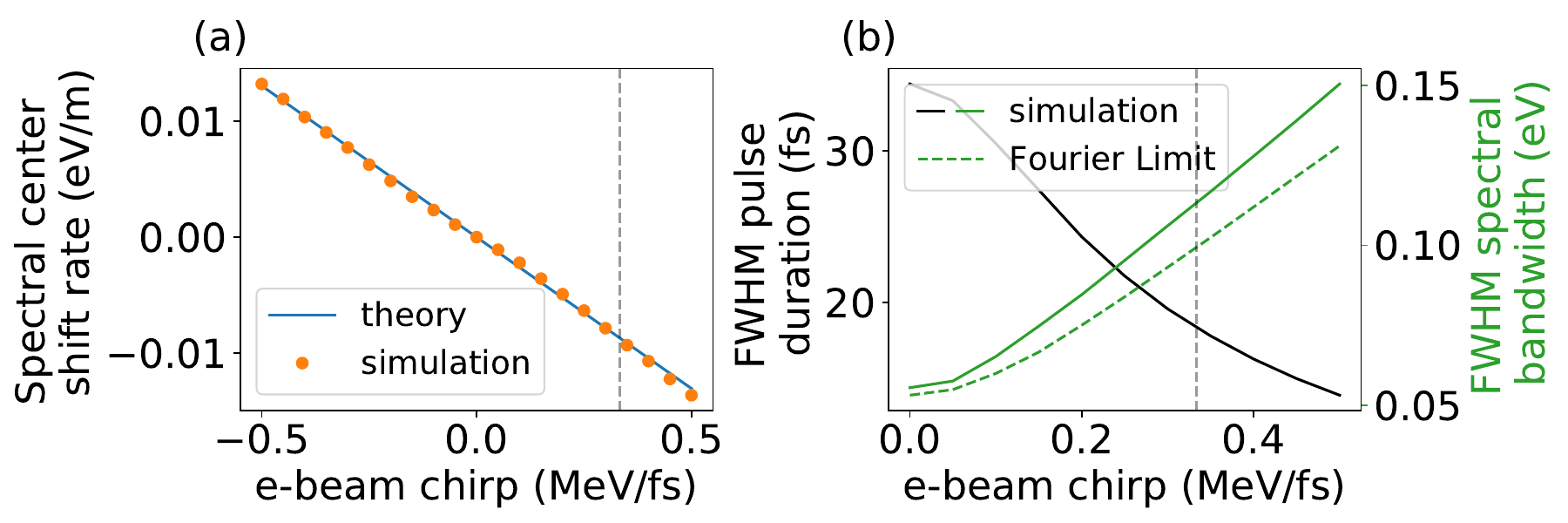}
   \caption{Spectral center shift and bandwidth broadening vs. linear e-beam energy chirp. (a) Spectral center shift rate as a function of initial e-beam chirp in the exponential gain regime. Theoretical prediction (blue line) is compared with 3D Genesis simulation (orange dots).  (b) Pulse length shortening (black) and bandwidth broadening (green) as a function of initial e-beam chirp at $z = 17m$ (near saturation). The fourier limit bandwidth of a Gaussian distribution is shown in the green dashed line. Simulations assume 200\, MW initial seed power. Beam and undulator parameters used are listed in Table 1 Case I. The vertical line indicates the e-beam chirp used in the Case I simulation.}
   \label{fig:theory_vs_sim}

\end{figure}

Fig.~\ref{fig:theory_vs_sim} (a) shows the spectral center shift as a function of initial energy chirp of the electron beam. The 3D FEL code GENESIS~\cite{REICHE1999243} is used to simulate seeded FEL process assuming an ideal 200\,MW initial seed and LCLS-II HE beam and undulator parameters as listed in Table I Case I. We linearly fit the spectrum center shift rate in the exponential growth regime and the results are consistent with theoretical predictions from Eq.2.  Fig.~\ref{fig:theory_vs_sim} (b) shows that the X-ray pulse is shorter with increasing energy chirp because the head and tail parts of the radiation are less amplified as their energies are further away from FEL resonant condition. The corresponding bandwidth is also increased but the time-bandwidth product is still very close to the Fourier-transform limit.

\section{Simulation with LCLS-II HE start-to-end beam}
The fast control of electron beam chirp can be achieved by using a normal-conducting (NC) RF cavity with a fast filling time, which has been proposed to control the electron energy compression and chirp on a shot-by-shot basis for the LCLS-II~\cite{nasr2016cw,zhang2023fast}. Here we show XRAFEL simluations with chirped electron beams from the start-to-end simulations from LCLS-II HE with a NC cavity.
\begin{figure}[htb]
   \centering
   \includegraphics*[width=0.5\columnwidth]{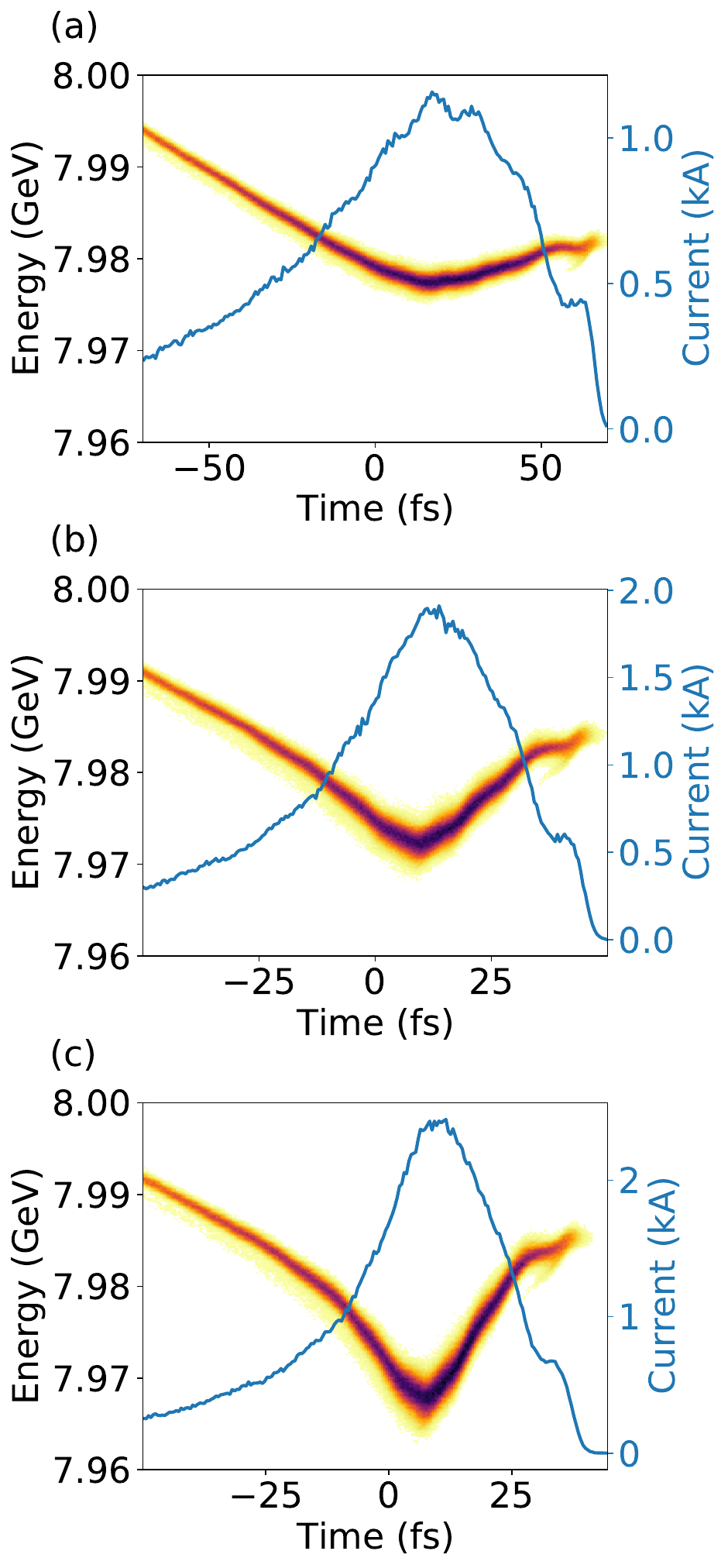}
   \caption{The longitudinal phase space and
the current profile of electron beams at the entrance of hard-Xray undulator from LCLS-II HE start-to-end simulation with the normal-conducting RF voltage $V_c = $0.0\,V (a), 0.22\,V(b) and 0.3\,V(c) respectively. Head if the bunch is the right of the figure.}
   \label{fig:NC_beams}
\end{figure}

Fig.~\ref{fig:NC_beams} shows the longitudinal phase space and current profiles of the electron beams at the entrance of hard Xray undulator line with different NC cavity voltages. The longitudinal phase space has a ``V" shape due to the space charge effects, where the energy chirp of both the tail and head part of the beam  is increased with the NC cavity voltage. By selectively matching the twiss parameters and controlling the undulator tapering, we can suppress the lasing from all but the head part of the ``V" shape beam. The beam chirper can introduce slight changes in the overall twiss parameters of the beam as the beam current varies, but it does not significantly affect the lasing performance along the beam. 

\begin{figure}[htb]
   \centering
   \includegraphics*[width=0.5\columnwidth]{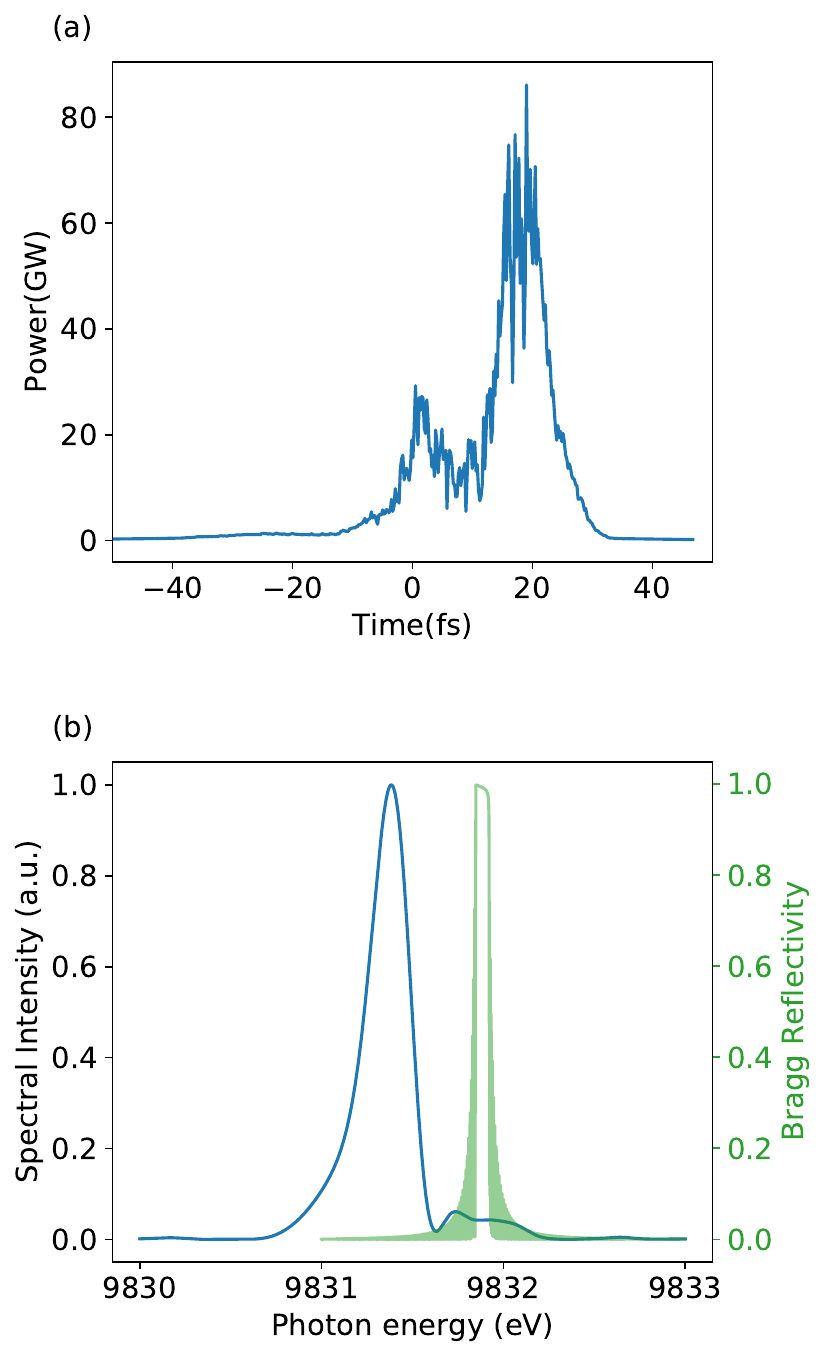}
   \caption{Power profile (a) and Spectrum (b) at the end of 124.8m undulator line. }
   \label{fig:power_spec}
\end{figure}

We use the chirped electron beam electron beams with $V_c = 0.22 V$ (as shown in Fig.~\ref{fig:NC_beams}(b)) for a GENESIS simulation. Fig.~\ref{fig:power_spec} shows the power profile and on-axis spectrum at the end of 124.8 undulators. By matching the twiss parameters of head part of the electron beam to the undulator line and optimizing the tapering, lasing from the tail part of the electron beam is suppressed. As a result, in main peak of the spectrum is red-shifted from the Bragg window.

\begin{figure}[htb]
   \centering
   \includegraphics*[width=0.6\columnwidth]{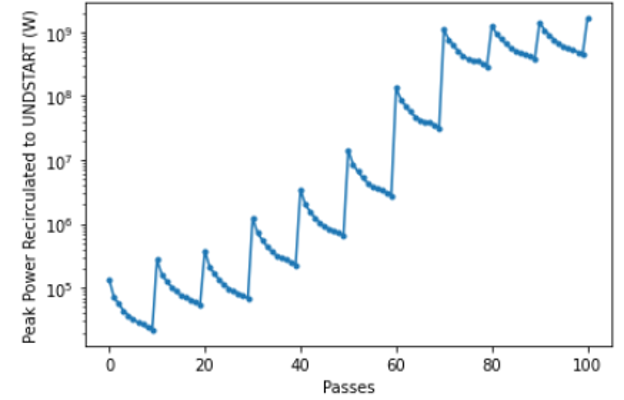}
   \caption{Intra-cavity power evolution starting from SASE as a function of recirulcation passes in the cavity, with start-to-end LCLS-II HE electron beams. }
   \label{fig:power_evolve}
\end{figure}

Figure~\ref{fig:power_evolve} shows an example of intra-cavity power build-up with start-to-end LCLS-II HE electron beams as shown in Fig.~\ref{fig:NC_beams} (b). Undulator and cavity parameters are the same as Case I. Electron beams on each shot are chirped by a NC cavity. After the first eight electron beams, the seed power at the entrance of the undulator reaches 200\,MW.

\section{ELECTRON ENERGY CHIRP JITTER}

\begin{figure}[!htb]
   \centering
   \includegraphics*[width=0.86\columnwidth]{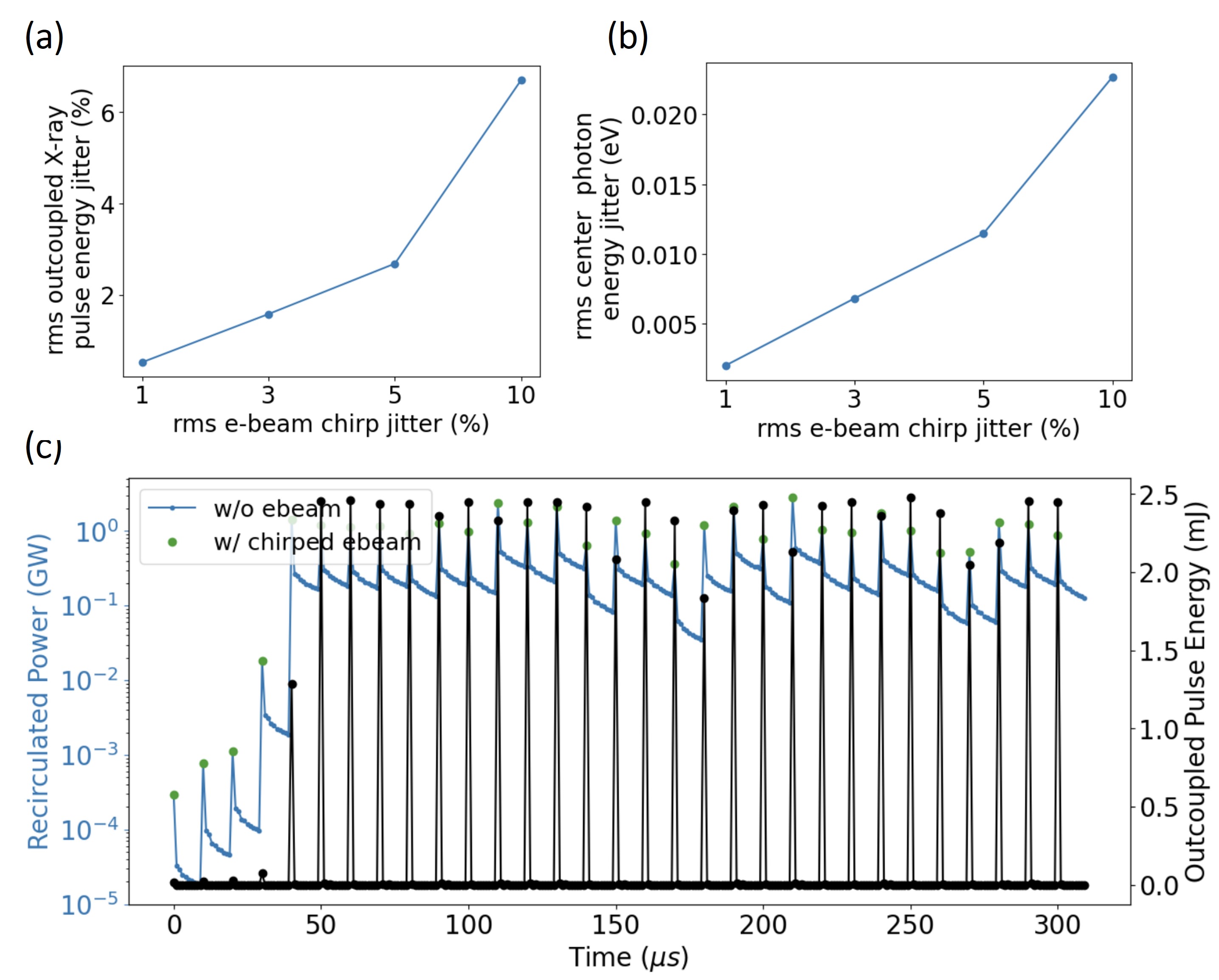}
   \caption{Simulation of electron-beam-based Q-switching scheme with electron energy chirp jittering. (a) (b) Stability of the outcoupled pulse as a function of e-beam chirp jitter. (c) The intra-cavity power buildup process with 10\% rms energy chirp jittering. }
   \label{fig:jitter}
\end{figure}

Here we discuss how the stability of the e-beam chirp will affect the stability of the FEL output. The FEL frequency shift has a linear dependence on the energy chirp of the electron beam. The narrow reflectivity bandwidth of the Bragg mirrors requires only a small shift in the center frequency to outcouple the radiation. By maintaining the energy chirp jitter at a few percent level, the resulting frequency jitter will be minimal.  Take LCLS-II HE for an example. The energy chirp of the electron beam originates from the space charge and the resistive-wall wakefields, which is related to the final beam current. The beam current jitter of LCLS-II HE  will be less than  4\% rms\cite{LCLSII_design_report}, resulting in an electron beam chirp jitter at the same level. Figure~\ref{fig:jitter} represents the simulation results with different levels of beam energy chirp jitter. As shown in Fig.~\ref{fig:jitter}(b), 4\% of chirp jitter corresponds to the center photon energy jitter  less than 0.01\,eV.

 The intra-cavity power build-up process remains unaffected by the e-beam chirp jitter, as indicated in Fig.~\ref{fig:jitter}. After reaching a steady state, the recirculated seed power is rather sensitive to the electron beam jitter since it comes from the power in the spectral tail. However, the FEL output power remains relatively stable as the FEL is in deep saturation regime, as indicated in Fig.~\ref{fig:jitter} (a) and (c).

\section{Cavity misalignment}

Cavity misalignment under these configurations may pose a challenge, as there are multiple round trips of X-rays within the cavity before interacting with the electron beams. However, the presence of refractive lenses in the cavity cause misalignment-induced trajectory errors only to oscillate rather than accumulate from one pass to the next~\cite{Qi2022}. This is particularly the case in a confocal cavity, as Ref.~\cite{Qi2022} also demonstrates that the spatial and angular deviation from mirror misalignment can be canceled after an even number of recirculations. Additionally, XRAFEL's high gain per pass provides optical guiding to the X-ray trajectories. Our numerical studies on cavity misalignment indicate robust power buildup for up to a few hundred nrad of rms angular error on the mirrors, which is within the state-of-art optomechanical tolerance.

%\bibliographystyle{IEEEtran}
%\bibliography{ref}% Produces the bibliography via BibTeX.

% Generated by IEEEtran.bst, version: 1.14 (2015/08/26)